\documentclass[prl,amsmath,amssymb,twocolumn,showpacs,floatfix]{revtex4}
\usepackage{graphicx}% Include figure files
\usepackage{dcolumn}% Align table columns on decimal point
\usepackage{bm}% bold math

\begin{document}

%%%%%%%%%%%%%%%%%%%%%%%%%%%%%%%%%%%%%%%%%%%%%%%%%%%
\title{Magnetism and Disorder Effects on $\mu$SR Measurements of the 
Magnetic Penetration Depth in Iron-Based Superconductors}

\author{J.E.~Sonier,$^{1,2}$ W.~Huang,$^1$ C.V.~Kaiser,$^1$, C.~Cochrane,$^1$ V.~Pacradouni,$^1$ 
S.A. Sabok-Sayr,$^1$, M.D.~Lumsden$^3$, B.C.~Sales$^3$, M.A.~McGuire$^3$, A.S.~Sefat$^3$ and D.~Mandrus$^{3,4}$}

\affiliation{$^1$Department of Physics, Simon Fraser University, Burnaby, British Columbia V5A 1S6, Canada \\
$^2$Canadian Institute for Advanced Research, Toronto, Canada \\
$^3$Oak Ridge National Laboratory, Oak Ridge, Tennessee 37831, USA \\
$^4$Department of Materials Science and Engineering, University of Tennessee, Knoxville, Tennessee 37996, USA}

\date{\today}
%%%%%%%%%%%%%%%%%%%%%%%%%%%%%%%%%%%%%%%%%%%%%%%%%%%%%%%

\begin{abstract}
It is shown that attempts to accurately deduce the magnetic penetration depth $\lambda$
of overdoped BaFe$_{1.82}$Co$_{0.18}$As$_2$ single crystals by transverse-field muon spin rotation 
(TF-$\mu$SR) are thwarted by field-induced magnetic order and strong vortex-lattice disorder.
We explain how substantial deviations from the magnetic field distribution of a 
nearly perfect vortex lattice by one or both of these factors is also significant for other
iron-based superconductors, and this introduces considerable 
uncertainty in the values of $\lambda$ obtained by TF-$\mu$SR.      
\end{abstract}

\maketitle
TF-$\mu$SR is routinely used to determine the magnetic penetration
depth $\lambda$ of type-II superconductors in the vortex state for the purpose of
obtaining indirect information on the energy gap structure \cite{Sonier:00}. The
magnetic field distribution $n(B)$ in the sample is determined by detecting the decay positrons from 
implanted positive muons that locally probe the internal fields, and
$\lambda$ is subsequently determined by modeling the contribution of the
vortex lattice (VL) to $n(B)$. However, even in conventional superconductors the
VL contribution is not known a priori, and one must rely on phenomenological models to deduce
what is really an ``effective'' penetration depth $\tilde{\lambda}$. One reason for this
is that only cumbersome microscopic theories account for the effects of low-energy excitations on $n(B)$ \cite{Ichioka:99}. Extrapolating low-temperature
measurements of $\tilde{\lambda}$ to zero field to eliminate the effects of intervortex
transfer of quasiparticles, as well as nonlocal and nonlinear effects, has been
demonstrated to be an accurate way of determining the 
``true'' magnetic penetration depth $\lambda$ \cite{Sonier:07a,Khasanov:09}.
Yet an underlying assumption is always that the VL
is highly ordered and that other contributions to $n(B)$ are relatively minor. 
The purpose of this Letter is to point out that this is not the case in many of the recently
discovered iron-based superconductors, making a reliable determination of $\lambda$
by TF-$\mu$SR extremely difficult.

Here we report on representative TF-$\mu$SR measurements of 
BaFe$_{1.82}$Co$_{0.18}$As$_2$ ($T_c \! = \! 21$~K) single crystals grown from a FeAs
flux, as described elsewhere \cite{Sefat:08}. High-statistics TF-$\mu$SR spectra 
of 20 million muon decay events were collected in
magnetic fields $H \! = \! 0.02$~T to 0.5 T applied {\it transverse} to the initial 
muon spin polarization $P(t \! = \! 0)$, and parallel to the $c$-axis of the crystals. 
The TF-$\mu$SR signal is the time evolution of the muon spin 
polarization, and is related to $n(B)$ as follows 
\begin{equation}
P(t) = \int_0^{\infty} n(B) \exp(i \gamma_\mu B t) dB \, ,
\label{eq:polarization}  
\end{equation}
where $\gamma_\mu$ is the muon 
gyromagnetic ratio. Generally, the TF-$\mu$SR signal is fit in the time domain, 
with the inverse Fourier transform or ``TF-$\mu$SR line shape'' given by
\begin{equation}
n(B) = \int_0^{\infty} P(t) \exp(-i \gamma_\mu B t) dt \, ,
\label{eq:distribution}  
\end{equation}
providing a visual approximation of the internal field distribution. 
The field distribution of a perfectly ordered VL is
characterized by sharp cutoffs at the minimum and maximum values of $B({\bf r})$, and
a sharp peak at the saddle-point value of $B({\bf r})$ \cite{Sonier:00}. These features 
are not observed in polycrystalline samples, where the 
orientation of the crystal lattice varies with respect to $H$, but
are observed in single crystals when a highly-ordered VL exists and other
contributions to $n(B)$ are relatively minor.   

\begin{figure}
\centering
\includegraphics[width=9.0cm]{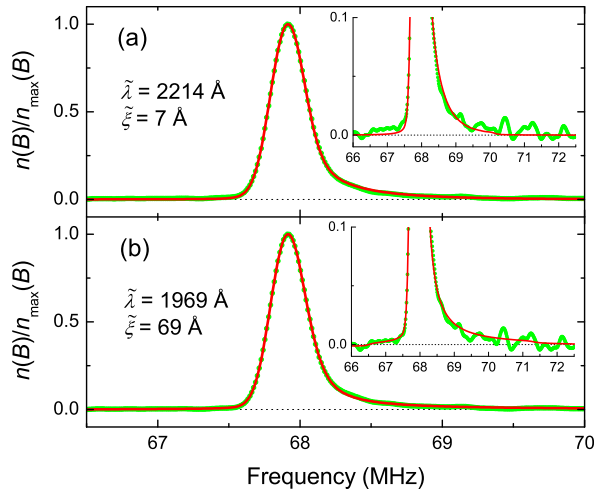}
\caption{(Color online) TF-$\mu$SR line shape of BaFe$_{1.82}$Co$_{0.18}$As$_2$
at $H \! = \! 0.5$~T and $T \! = \! 3.9$~K (green circles).
(a) The red curve is the Fourier transform of a fit in the time domain assuming $B({\bf r})$
of the ideal VL is given by Eq.~(\ref{eq:GLfield}). In addition to
the indicated values of $\tilde{\lambda}$ and $\tilde{\xi}$, the fit yields
$\sigma \! = \! 0.265$~$\mu$s$^{-1}$ and a PM shift of 8.6~G.             
(b) Fourier transform of a fit in the time domain that 
assumes the model of field-induced AF order described in the main text (red curve). 
Some of the fit parameters are shown in Fig.~\ref{fig3}.}
\label{fig1}
\end{figure}

We have attempted to fit the TF-$\mu$SR spectra to a theoretical polarization 
function $P(t)$ that has been successfully applied to a wide variety of type-II
superconductors, and utilized in some of the experiments on
iron-based superconductors. The spatial variation of the field,
from which $n(B)$ is calculated, is modeled by the analytical 
Ginzburg-Landau (GL) function \cite{Sonier:00}
\begin{equation}
B({\bf r}) = B_0 (1-b^4) \sum_{ {\bf G}}
\frac{e^{-i {\bf G} \cdot {\bf r}} \, \, u \, K_1(u)}{\tilde{\lambda}^2 G^2} \, ,
\label{eq:GLfield}
\end{equation}
where {\bf G} are the reciprocal lattice vectors of an hexagonal VL, $b \! = \! B / B_{c2}$ 
is the reduced field, $B_0$ is the average internal magnetic field, $K_1(u)$ is a modified 
Bessel function, and $u^2 \! = \! 2 \tilde{\xi}^2 G^2 (1 + b^4)[1-2b(1 - b)^2]$. As
explained later, $P(t)$ is multiplied by a Gaussian depolarization function 
$\exp(-\sigma^2 t ^2)$ to account for the effects of nuclear dipolar fields and
frozen random disorder. 
We stress that the fitting parameters $\tilde{\xi}$ and $\tilde{\lambda}$ can deviate substantially
from the ``true'' coherence length and magnetic penetration depth if other 
contributions to $n(B)$ are significant.  
An important feature of Eq.~(\ref{eq:GLfield}) is that it accounts for the 
finite size of the vortex cores, by generating a ``high-field'' cutoff in
$n(B)$. The GL coherence length $\xi_{ab} \! \sim \! 26$~\AA~ calculated
from the upper critical field $H_{c2} \sim \! 50$~T of 
BaFe$_{1.84}$Co$_{0.16}$As$_2$ with $H \! \parallel \! c$ \cite{Kano:09},
represents a lower limit for the vortex core radius \cite{Sonier:07a}. 
The core size can be much larger if there are spatially extended quasiparticle
core states associated with either the existence of a second 
smaller superconducting gap \cite{Callaghan:05} or a single anisotropic gap
\cite{Sonier:09}. Yet fits of the TF-$\mu$SR spectra of BaFe$_{1.82}$Co$_{0.18}$As$_2$ 
using Eq.~~(\ref{eq:GLfield}), show no sensitivity to the vortex cores at 
any field and converge with values of $\tilde{\xi}$ approaching zero.
Fig.~\ref{fig1} shows that even at 0.5~T where the vortex density 
is highest, a high-field cutoff is not discernible in the TF-$\mu$SR line shape.
We next discuss reasons for this insensitivity to the vortex cores.
    
\begin{figure}
\centering
\includegraphics[width=9.0cm]{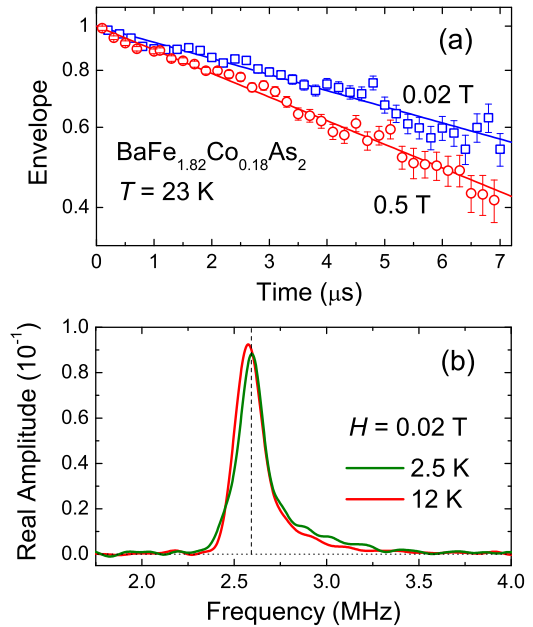}
\caption{(Color online) (a) Envelopes of TF-$\mu$SR spectra of BaFe$_{1.82}$Co$_{0.18}$As$_2$
in the normal state at $T \! = \! 23K$. The solid curves are fits to a single exponential
relaxation function $G(t) \! = \! \exp(-\Lambda t)$, yielding 
$\Lambda \! = \! 0.081 \! \pm \! 0.003$~$\mu$s$^{-1}$ and
$\Lambda \! = \! 0.119 \! \pm \! 0.003$~$\mu$s$^{-1}$ at $H \! = \! 0.02$~T and 
$H \! = \! 0.5$~T, respectively. (b) TF-$\mu$SR line shapes of BaFe$_{1.82}$Co$_{0.18}$As$_2$
in the superconducting state at $H \! = \! 0.02$~T. The dashed vertical line indicates
the magnitude of the applied field $H$.}
\label{fig2}
\end{figure}

\vspace{3mm}
\noindent {\it Magnetism}---The effective field {\bf B}$_{\mu}$ experienced by
the muon is a vector sum of various contributions that may be static or fluctuating
in time. With correlation times generally much longer than the muon life time,
the nuclear moments constitute a {\it dense} static moment system that cause
a Gaussian-like depolarization of the TF-$\mu$SR spectrum. Yet as shown in 
Fig.~\ref{fig2}(a), BaFe$_{1.82}$Co$_{0.18}$As$_2$ 
exhibits an exponential depolarization above $T_c$ that is typical of dilute 
or fast fluctuating electronic moments \cite{Uemura:85}. 
The latter interpretation is consistent with the observation of
a paramagnetic (PM) shift of the average internal field $\langle${\bf B}$_{\mu}\rangle$  
sensed by the muons below $T_c$.
This is evident in Fig.~\ref{fig2}(b), where we show representative Fourier transforms 
of $P(t)$ at $H \! = \! 0.02$~T. Instead of the expected diamagnetic shift imposed by the
superconducting state, $\langle${\bf B}$_{\mu}\rangle$ exceeds $H$.
The magnitude of the PM shift increases with increasing $H$ and/or decreasing $T$. 

The occurrence of a PM shift in the superconducting state of BaFe$_{2-x}$Co$_{x}$As$_2$ and 
SrFe$_{2-x}$Co$_{x}$As$_2$ has been reported by others \cite{Khasanov:09b,Williams:10}, and
implies an enhancement of $\langle${\bf B}$_{\mu}\rangle$ from magnetic order
occupying a large volume of the sample. 
Magnetic order exists in underdoped samples at $H \! = \! 0$ \cite{Marsik:10},
and is apparently induced in overdoped samples by the applied field. 
Yet the effects of magnetism on the line width and functional form of 
$n(B)$ have not been considered. A strong relaxation of the 
TF-$\mu$SR signal occurs even in long-range magnetically ordered systems, and
with decreasing temperature there must be an increased broadening of $n(B)$
associated with the growth of the correlation time for spin fluctuations. 

\begin{figure}
\centering
\includegraphics[width=9.0cm]{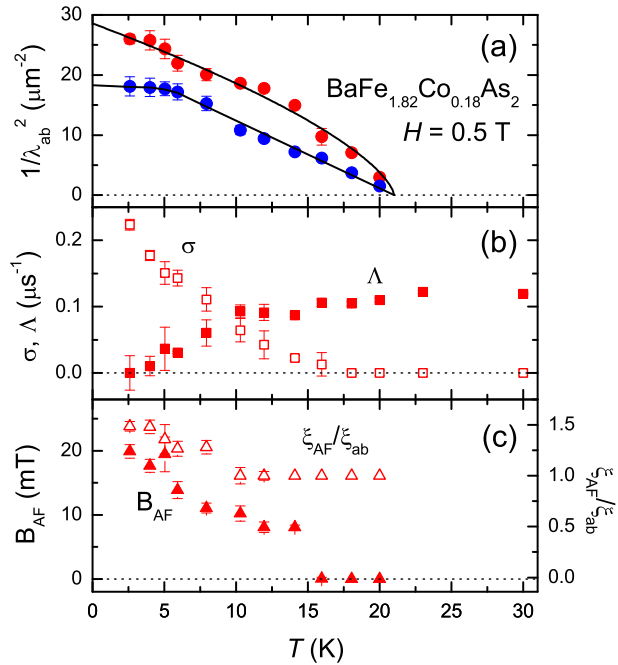}
\caption{(Color online) Results of fits of the TF-$\mu$SR time spectra of
BaFe$_{1.82}$Co$_{0.18}$As$_2$ at $H \! = \! 0.5$~T, assuming the 
model of magnetic order described in the main text. 
Temperature dependence of (a) $1/\tilde{\lambda}^2$, (b) the depolarization rates
$\sigma$ and $\Lambda$, and (c) the amplitude of the additional AF field in the 
vortex-core region.
Also shown in (a) are results of fits without magnetic order and $\tilde{\xi} \! = \! 5$~\AA
(blue circles).}
\label{fig3}
\end{figure} 

Accounting for such magnetism is non-trivial because of the 
spatially-varying superconducting order parameter and
the likelihood that the field-induced magnetism occurs in a nematic phase 
\cite{Chu:10}. Even so we have achieved excellent fits of the TF-$\mu$SR spectra of 
BaFe$_{1.82}$Co$_{0.18}$As$_2$ to polarization functions that incorporate 
enhanced magnetism in the vortex core region ({\it e.g.} commensurate spin-density wave, 
ferromagnetism, spin-glass), where superconductivity is suppressed. 
Here we describe typical results for one form of magnetism:
To account for line broadening from magnetism at higher temperatures,
$P(t)$ is multiplied by an exponential depolarization function $\exp(-\Lambda t)$, 
as observed above $T_c$. In addition, enhanced magnetic order in the vortex cores 
is modeled by adding the following term to Eq.~(\ref{eq:GLfield}) 
\begin{equation}
B_{\rm AF}({\bf r}) = B_{\rm AF} e^{-\frac{1}{2}(r / \xi_{\rm AF})^2} 
\sum_{ {\bf K}} \left( e^{-i {\bf K} \cdot {\bf r}} - e^{-i {\bf K} \cdot {\bf r}^\prime}
\right) \, .
\label{eq:AFfield}
\end{equation}
The {\bf K} sum is the reciprocal lattice of an antiferromagnetic (AF) square iron 
sublattice of spacing $a \! = \! 2.8$~\AA, $B_{\rm AF}$ is the field amplitude, 
$\xi_{\rm AF}$ governs the radial decay of the field amplitude from the core center, and 
{\bf r} and {\bf r}$^\prime$  
are the position vectors for `up' and `down' spins, respectively.
This kind of magnetic order has the effect of smearing the 
high-field cutoff, and can even introduce a low-field tail in $n(B)$ \cite{Sonier:07b}.  
As indicated by the value of $\tilde{\xi}$ in Fig.~\ref{fig1}(b), fits to this model 
are sensitive to the vortex cores. With decreasing temperature, the magnetism-induced
relaxation evolves from exponential to Gaussian, and the magnetic order 
in the vortex cores is enhanced. Consistent with the results Ref.~\cite{Williams:10}, 
fits of TF-$\mu$SR spectra of overdoped BaFe$_{2-x}$Co$_{x}$As$_2$
to a model that does not include magnetism and is insensitive to the vortex cores 
({\it i.e.} $\tilde{\xi}$ fixed to 5~\AA) yield an unusual linear temperature of 
$1/\tilde{\lambda}^2$ immediately below $T_c$, and a saturation of $\tilde{\lambda}$ 
at low $T$. In contrast, fits assuming magnetic order exhibit a linear temperature 
dependence well below $T_c$ that is suggestive of gap nodes. 
However, these results simply demonstrate the ambiguity in modeling such data. 
Without knowledge of the precise form of the magnetism, our model 
cannot be deemed rigorously valid. Furthermore, as we explain next, VL disorder
is a serious concern.

\begin{figure*}
\centering
\includegraphics[width=19.0cm]{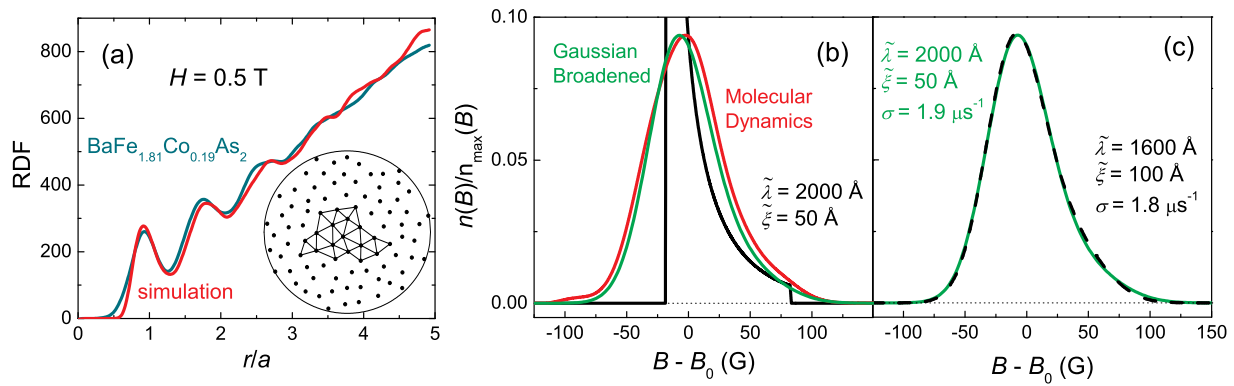}
\caption{(Color online) (a) Radial distribution function (RDF) of 
BaFe$_{1.81}$Co$_{0.19}$As$_2$ at $H \! = \! 0.5$~T from Ref.~\cite{Inosov:10} 
and of the disordered VL shown in the lower right generated by
molecular dynamics (MD). Note 5000 vortices were used in the MD simulation.
The horizontal scale is normalized with respect to the intervortex spacing 
$a \! = \! 691$~\AA~ of the perfect hexagonal VL. (b) Theoretical simulations 
of the TF-$\mu$SR line shape of the perfect
VL (black curve) and of the disordered VL (red curve) corresponding to the RDF shown in (a).
The green curve is the line shape of the perfect VL convoluted by a Gaussian distribution of
fields, corresponding to $\sigma \! = \! 1.9$~$\mu$s$^{-1}$.  
All three simulations assume $\tilde{\lambda} \! = \! 2000$~\AA and 
$\tilde{\xi} \! = \! 50$~\AA. 
(c) Same disordered line shape shown in (b) and a Gaussian broadened
ideal line shape with $\tilde{\lambda} \! = \! 1600$~\AA,
$\tilde{\xi} \! = \! 100$~\AA, and $\sigma \! = \! 1.8$~$\mu$s$^{-1}$.
The heights of the line shapes in (b) and (c) are normalized with repect to the height
$n_{\rm max}(B)$ of the ideal line shape.}  
\label{fig4}
\end{figure*}

\vspace{3mm}
\noindent {\it Disorder}---Thus far TF-$\mu$SR has been applied to iron-based 
superconductors under the assumption that one is probing a fairly well-ordered 
hexagonal VL. Yet to date this has been observed
only in KFe$_2$As$_2$ \cite{Kawano:10}. 
Vortex imaging experiments on the $R$FeAs(O$_{1-x}$F$_x$), 
$A_{1-x}B_x$Fe$_2$As$_2$ and $A$Fe$_{2-x}$Co$_x$As$_2$ families 
all show a highly disordered VL indicative of strong bulk 
pinning \cite{Yin:09,Eskildsen:09a,Eskildsen:09b,Vinnikov:09,Inosov:10,Luan:10}.
In Fig.~\ref{fig4} we show the effect of such disorder on the ideal $n(B)$.
Using a radial distribution function closely resembling that observed in overdoped 
BaFe$_{1.81}$Co$_{0.19}$As$_2$ \cite{Inosov:10}, we have used
molecular dynamics to simulate $n(B)$ of the disordered VL.
Although the disordered line shape in Fig.~\ref{fig4}(b) is asymmetric, it is strongly 
smeared with a field variation greatly exceeding that of the perfect VL.   
 
Small perturbations of the VL by random pinning
can be handled by convoluting the ideal theoretical line shape with 
a Gaussian distribution of fields \cite{Brandt:88}. This causes a Gaussian depolarization
$\exp(- \sigma^2 t^2)$ of $P(t)$. But
for polycrystalline samples, $n(B)$ of the
perfect VL is nearly symmetric and cannot be isolated from a symmetric distribution
of disorder. Consequently, VL disorder has not been accounted for in TF-$\mu$SR studies of 
polycrystalline or powdered iron-based superconductors 
\cite{Luetkens:08,Khasanov:08,Carlo:09}. Given the severity of disorder in these materials
and no knowledge about how this disorder evolves with temperature or
doping, the accuracy of information deduced about $\lambda$ is questionable. 
Since disorder of rigid flux lines broaden $n(B)$,
such studies certainly underestimate $\lambda$.

While small perturbations of $B({\bf r})$
by vortex pinning may be accounted for in measurements on single crystals,  
a Gaussian convolution of the ideal $n(B)$ becomes increasingly 
inadequate as the degree of disorder is enhanced \cite{Menon:06}.
In Fig.~\ref{fig4}(b) we show that Gaussian broadening of the ideal line shape does not
precisely reproduce $n(B)$ of the disordered VL. More importantly, however, because
the large disorder-induced broadening smears out the high-field cutoff, 
the fitting parameters $\tilde{\lambda}$ and $\tilde{\xi}$ are ambiguous. 
This is illustrated in      
Fig.~\ref{fig4}(c), where a nearly identical Gaussian broadened line shape is
obtained for very different values of these parameters. Hence 
disorder introduces considerable uncertainty even in measurements on single crystals
\cite{Khasanov:09b,Williams:10,Aczel:08,Khasanov:09a,Goko:09,Williams:09}.

In summary, the effects of magnetic order and/or random frozen disorder of the VL 
in iron-based superconductors introduce considerable uncertainty in values of $\lambda$ 
obtained by TF-$\mu$SR.
Unfortunately, these effects cannot be modeled in a reliable way. Compounding the problem 
is a lack of information on how the magnetism and VL disorder evolve with temperature. 
Consequently, caution is warranted in drawing conclusions about the  
anisotropy of the superconducting gap in these materials from TF-$\mu$SR measurements.

We thank the staff of TRIUMF's 
Centre for Molecular and Materials Science for technical assistance. 
Work at TRIUMF was supported by NSERC of Canada, and at ORNL by the U.S. DOE,
Office of Basic Energy Sciences, Materials Science and Engineering Division.


\begin{thebibliography}{xx}

\bibitem{Sonier:00} J.E.~Sonier, J.H.~Brewer and R.F.~Kiefl, Rev.~Mod.~Phys. {\bf 72}, 769 (2000).

\bibitem{Ichioka:99} M.~Ichioka, A.~Hasegawa and K.~Machida, Phys.~Rev.~B {\bf 59}, 184 (1999).

\bibitem{Sonier:07a} J.E.~Sonier, Rep.~Prog.~Phys. {\bf 70}, 1717 (2007).

\bibitem{Khasanov:09} R.~Khasanov {\it et al.}, Phys.~Rev.~B {\bf 79}, 180507(R) (2009).

\bibitem{Sefat:08} A.S.~Sefat {\it et al.}, Phys.~Rev.~Lett. {\bf 101}, 117004 (2008).

\bibitem{Kano:09} M.~Kano {\it et al.}, J.~Phys.~Soc.~Jpn. {\bf 78}, 084719 (2009).

\bibitem{Callaghan:05} F.D.~Callaghan {\it et al.}, Phys.~Rev.~Lett. {\bf 95}, 197001 (2005).

\bibitem{Sonier:09} J.E.~Sonier {\it et al.}, Phys.~Rev.~Lett. {\bf 83}, 4156 (1999). 
  
\bibitem{Uemura:85} Y.J.~Uemura, T.~Yamazaki, D.R.~Harshman, M.~Senba and E.J.~Ansaldo,
Phys.~Rev.~B {\bf 31}, 546 (1985).

\bibitem{Marsik:10} P.~Marsik {\it et al.}, Phys.~Rev.~Lett. {\bf 105}, 057001 (2010).

\bibitem{Khasanov:09b} R.~Khasanov {\it et al.}, Phys.~Rev.~Lett. {\bf 103}, 067010 (2009).    

\bibitem{Williams:10} T.J.~Williams {\it et al.}, Phys.~Rev.~B {\bf 82}, 094512 (2010).

\bibitem{Chu:10} J.-H.~Chu {\it et al.}, Science {\bf 329}, 824 (2010).

\bibitem{Sonier:07b} J.E.~Sonier, Phys.~Rev.~B {\bf 76}, 064522 (2007).

\bibitem{Kawano:10} H.~Kawano-Furukawa {\it et al.}, arXiv:1005.4468.

\bibitem{Yin:09} Y.~Yin {\it et al.}, Phys.~Rev.~Lett. {\bf 102}, 097002 (2009). 

\bibitem{Eskildsen:09a} M.R.~Eskildsen {\it et al.}, Phys.~Rev.~B {\bf 79}, 100501(R) (2009).

\bibitem{Eskildsen:09b} M.R.~Eskildsen {\it et al.}, Physica~C {\bf 469}, 529 (2009).

\bibitem{Vinnikov:09} L.Ya.~Vinnikov {\it et al.} JETP~Lett. {\bf 90}, 299 (2009).

\bibitem{Inosov:10} D.S.~Inosov {\it et al.}, Phys.~Rev.~B {\bf 81}, 014513 (2010).

\bibitem{Luan:10} L.~Luan {\it et al.}, Phys.~Rev.~B {\bf 81}, 100501 (2010).

\bibitem{Brandt:88} E.H.~Brandt, Phys.~Rev.~B {\bf 37}, 2349 (1988).

\bibitem{Luetkens:08} H.~Luetkens {\it et al.}, Phys.~Rev.~Lett. {\bf 101}, 097009 (2008).

\bibitem{Khasanov:08} R.~Khasanov {\it et al.}, Phys.~Rev.~B {\bf 78}, 092506 (2008).

\bibitem{Carlo:09} J.P.~Carlo {\it et al.}, Phys.~Rev.~Lett. {\bf 102}, 087001 (2009).

\bibitem{Menon:06} G.I.~Menon {\it et al.}, Phys.~Rev.~Lett. {\bf 97}, 177004 (2006).

\bibitem{Aczel:08} A.A.~Aczel {\it et al.}, Phys.~Rev.~B {\bf 78}, 214503 (2008).

\bibitem{Khasanov:09a} R.~Khasanov {\it et al.}, Phys.~Rev.~Lett. {\bf 102}, 187005 (2009).

\bibitem{Goko:09} T.~Goko {\it et al.}, Phys.~Rev.~B {\bf 80}, 024508 (2009).

\bibitem{Williams:09} T.J.~Williams {\it et al.}, Phys.~Rev.~B {\bf 80}, 094501 (2009).

\end{thebibliography}
\end{document}